# Dynamics of single Au nanoparticles on graphene simultaneously in real- and diffraction space by time-series convergent beam electron diffraction

Sara Mustafi[1,2], Rongsheng Cai[3,4], Sam Sullivan-Allsop[3,4], Matthew Smith[3,4], Nicholas J. Clark[3,4], Matthew Lindley[3,4], Ding Peng[1], Kostya S. Novoselov[4,5,6,7], Sarah J. Haigh[3,4], Tatiana Latychevskaia[1,2,*]

[1]Paul Scherrer Institute, Forschungsstrasse 111, 5232 Villigen, Switzerland

[2]Physics Department, University of Zurich, Winterthurerstrasse 190, 8057 Zurich, Switzerland

[3]Department of Materials, University of Manchester, Oxford Road, Manchester, M13 9PL, UK

[4]National Graphene Institute, University of Manchester, Oxford Road, Manchester, M13 9PL, UK

[5]Department of Materials Science and Engineering, National University of Singapore, Singapore, 117575, Singapore

[6]Institute for Functional Intelligent Materials, National University of Singapore, 117544 Singapore

[7]Chongqing 2D Materials Institute, Liangjiang New Area, Chongqing, 400714, China

Corresponding author: tatiana.latychevskaia@psi.ch

## Abstract

Convergent beam electron diffraction (CBED) on two-dimensional materials allows simultaneous recording of the real-space image (tens of nanometers in size) and diffraction pattern of the same sample in one single-shot intensity measurement. In this study, we employ time-series CBED to visualize single Au nanoparticles deposited on graphene. The real-space image of the probed region, with the amount, size, and positions of single Au nanoparticles, is directly observed in the zero-order CBED disk, while the atomic arrangement of the Au nanoparticles is available from the intensity distributions in the higher-order CBED disks. From the time-series CBED patterns, the movement of a single Au nanoparticle with rotation up to 4° was recorded. We also observed facet diffraction lines – intense bright lines formed between the CBED disks of the Au nanoparticle, which we explain by diffraction at the Au nanoparticle's facets. This work showcases CBED as a useful technique for studying adsorbates on graphene using Au nanoparticles as a model platform, and paves the way for future studies of different objects deposited on graphene.

# 1. Introduction

Transmission electron microscopy (TEM) is a practical tool for imaging the structure of materials at atomic resolution in real space by using aberration-corrected transmission electron microscopes (TEMs) [1-4] or in diffraction space by applying lensless imaging techniques such as coherent diffraction imaging (CDI) [5, 6] and ptychography [7]. Each of these modes has particular usage and provides specific information. At the same time, it would be extremely beneficial to combine the possibilities of real and diffraction space imaging into one technique. By probing a two-dimensional (2D) crystal with a convergent electron wavefront, the real-space image (tens of nanometers in size) and diffraction pattern of the same region are acquired in one single-shot CBED pattern [8]. CBED on 2D materials was previously demonstrated for single-layer graphene, twisted bilayers, and for some residuals found on top of graphene. The goal of this study is to investigate what information is provided by the CBED when well-studied test objects of known structure – Au nanoparticles – are placed on top of graphene. A time-series convergent beam electron diffraction (CBED) is employed to study dynamical processes in single Au nanoparticles.

## 1.1 Convergent beam electron diffraction (CBED) of 2D crystals

Conventional CBED has been applied for decades to study the structure of three-dimensional (3D) crystals [9] and provide information about lattice parameters [10], structure factors [11], sample thickness [12], and defects [13], with the structure reconstruction done by performing simulations that match the experimental observations. Recently, CBED has been demonstrated for the imaging of 2D crystals [8, 14, 15]; an experimental arrangement is shown in Fig. 1a. CBED of a 2D crystal (graphene, hexagonal boron nitride, etc.) allows direct imaging of the real-space distribution of the sample in the probed region and its diffraction pattern in one single-shot intensity measurement. The convergence semi-angle is selected so that the CBED disks do not overlap, unlike in electron ptychography (or 4D-STEM), where a large convergence semi-angle ensures that the CBED disks overlap – a necessary condition for phase recovery [16]. In the case of perfectly flat, clean graphene, the CBED disks of uniform intensity cover the entire detector area, thus providing a reference background that coherently interferes with the wave scattered by the object deposited on graphene. Due to the finite convergence of the probing beam, each Bragg peak is transformed into a disk of diameter proportional to the diameter of the probed area $D=2\,|\Delta f|\tan\alpha$, where $\Delta f$ is the defocus and $\alpha$ is the convergence semi-angle. The intensity distribution within a CBED disk is directly related to the real-space arrangement of atoms in the probed region. Any intensity fluctuation is a signature of lattice corrugation, defect, or an adsorbate. Each CBED disk exhibits a unique interference pattern and can be analyzed independently. Intensity distribution in a CBED disk is in fact an in-line hologram, and the sample distribution can be reconstructed from the intensity distribution by applying reconstruction algorithms adapted from holography [17]. Previously, nano-sized objects deposited on the graphene surface were reconstructed from the zero-order CBED disk by applying in-line holography reconstruction algorithms and features as small as 8 Å were resolved [18]. The first- and higher-order CBED disks contain atomic-resolution information about the sample. In-plane and out-of-plane atomic displacements (corrugations) [8, 18, 19], atomic defects [20], the projected potentials of individual atoms in graphene [21], the number of layers [22] and the interlayer distance in van der Waals structures [23, 24] were recovered from the CBED patterns of 2D crystals.

When Au nanoparticles are deposited on the top of graphene, the amount, size, and positions of Au nanoparticles in the probed region are directly seen in the zero-order CBED disk – the in-line hologram of the sample (Fig. 1a, b). Such information is not accessible from a conventional diffraction pattern of Au nanoparticles. The zero-order peak is also usually blocked by a beamstop in diffraction patterns. In addition, the CBED pattern captures higher-order CBED disks of the probed Au nanoparticle, where the position and intensity of the disks carry the information about the orientation of the Au nanoparticle and its atomic arrangement. Thus, a single CBED pattern simultaneously provides information about the exact location of an Au nanoparticle on the graphene support and high-resolution information about the crystalline structure of the Au nanoparticle. The structural metamorphosis of an Au nanoparticle can be monitored by acquiring a time series of CBED patterns. The main motivation of this study is to demonstrate CBED of 2D crystals as a powerful technique for imaging of objects deposited on graphene by using Au nanoparticles as test objects.

## 1.2 Au nanoparticles

Nanometer-sized Au nanoparticles can be found in noncrystalline icosahedral (Ih) and decahedral (Dh) and crystalline cuboctahedral (Ch) face-centred cubic (FCC) structures. The Ih structure exhibits high internal strain and is favorable for very small sizes (2 nm or less) where the surface energy minimization exceeds the strain contribution that is proportional to the particle's volume [25]. The Dh structure exhibits less internal strain and thus becomes more favorable for larger sizes. The FCC structure does not have an internal strain and thus is the dominant structure for particles of large size and for bulk metallic gold.

The structure and behavior of Au nanoparticles have been extensively studied by electron microscopy at atomic resolution [26, 27]. The temperature increase due to electron beam irradiation has been measured to be within 20 – 50°C [28] – insufficient to cause Au nanoparticle melting. Upon electron beam irradiation, small (less than 5 nm in diameter) Au nanoparticles exhibit fast structural transformations between Ih, Dh, and Ch FCC isomers [4, 25, 29-37]. Room-temperature TEM studies show that the Dh is the preferred structure, followed by the Ih structure that is less stable and shows transition to Dh structures [4, 38]. Recently, it was suggested that the experimentally observed structural transformations upon exposure to 200 keV electrons can be induced by the relaxation of collective electronic (plasmon) excitations formed in deposited metal clusters due to the interaction with low-energy secondary electrons emitted from a substrate [39]. Small Au nanoparticles have a melting point that is smaller than that of bulk gold – 1064°C [40]. The melting temperature of small nanoparticles in Pawlow's model is inversely proportional to the particle's radius as $-1/r$ and increases as the particle's radius increases [41]. The ($-1/r$) dependency was experimentally confirmed by atomic-resolution TEM imaging of individual Au nanoparticles melting at temperatures increasing from 20° to 1000°C [33], where the melting temperature was measured to be 800 - 900°C for nanoparticles of 2.5 nm in diameter. Structural and space transformations of individual Au nanoparticles at the temperatures close to the melting temperature, with the majority of transformations to Dh structures, have been observed by atomic-resolution TEM [42, 43].

Dynamical behavior of Au nanoparticles [44] – movements or rotations – requires much less energy than structural transformation. For a 5 nm in diameter Au nanoparticle, the energy of a lateral movement by 1 nm in 1 ns is 3.942 meV, and the energy of a rotation by 5 degrees in 1 ns is 0.075 meV, giving a high probability of these movements at room temperature $\sim\exp(-E/k_BT)$ = 0.858 and 0.997, respectively; $k_BT$ = 25.6 meV. Thus, this dynamical behavior of Au nanoparticles can be expected even without the presence of an electron beam. Wang et al. observed in their aberration-corrected TEM studies of $Au_{923}$ nanoparticles that the Ih structures were not stable and converted into Dh or FCC structures; the conversion can happen quickly even during the beam alignment [45].

TEM studies show that all metals, as singular atoms or atom aggregates, reside in the omni-present hydrocarbon surface contamination rather than on pristine graphene [46]. Au atoms can form 2D clusters that undergo rotation to achieve matching of planes in the Au <110> and graphitic [001] zones [47]. The binding energies calculated by density functional theory showed that a 34-atom Au cluster can be anchored on single-vacancy defective-graphene, tungsten diselenide, and thiophenol functionalized graphene, but not on graphene or hexagonal boron-nitride [48]. Theoretical calculations suggest that the presence of the graphene vacancy enhances the bonding of the nanoparticles to the graphene support [49, 50].

## 2. Theory and simulations
### 2.1 CBED patterns of Au nanoparticles on graphene

For a single Au nanoparticle residing on graphene, three signature signals are present in the CBED pattern: CBED disks from graphene, CBED disks of the Au nanoparticle, and what we refer to as "shadow images" of the Au nanoparticle. The CBED pattern of graphene is composed of six-fold-symmetry-arranged CBED disks (Fig. 1b, c), whose positions and intensity remain unchanged when the graphene sheet is laterally shifting. The CBED pattern of a single Au nanoparticle has almost the same appearance as is expected for the Au nanoparticle's parallel beam electron diffraction pattern: the diameter of the Au nanoparticle's CBED disks is given by the diameter of the Au nanoparticle and thus is so small that these disks can be referred to as "peaks" (Fig. 1b, d); however, to avoid any confusion, we will refer to them as "disks". There are important differences between a parallel beam electron diffraction pattern and CBED pattern for an Au nanoparticle. In the parallel beam electron diffraction pattern of an Au nanoparticle, the zero-order peak is the most intense and bright peak, and the diffraction pattern is centred at the optical axis. In the CBED pattern, the zero-order CBED disk is a disk of uniform intensity created by the probing beam with a darker region (shadow image) of Au nanoparticle (Fig. 1a, b, and d). The Au nanoparticle's diffraction pattern is centred at the Au nanoparticle's location, that is, at the shadow image's position in the zero-order CBED disk. The relative positions of the Au nanoparticle's CBED disks are the same as the positions of the parallel beam electron diffraction peaks. When the Au nanoparticle is laterally shifting, its CBED pattern is also shifting.

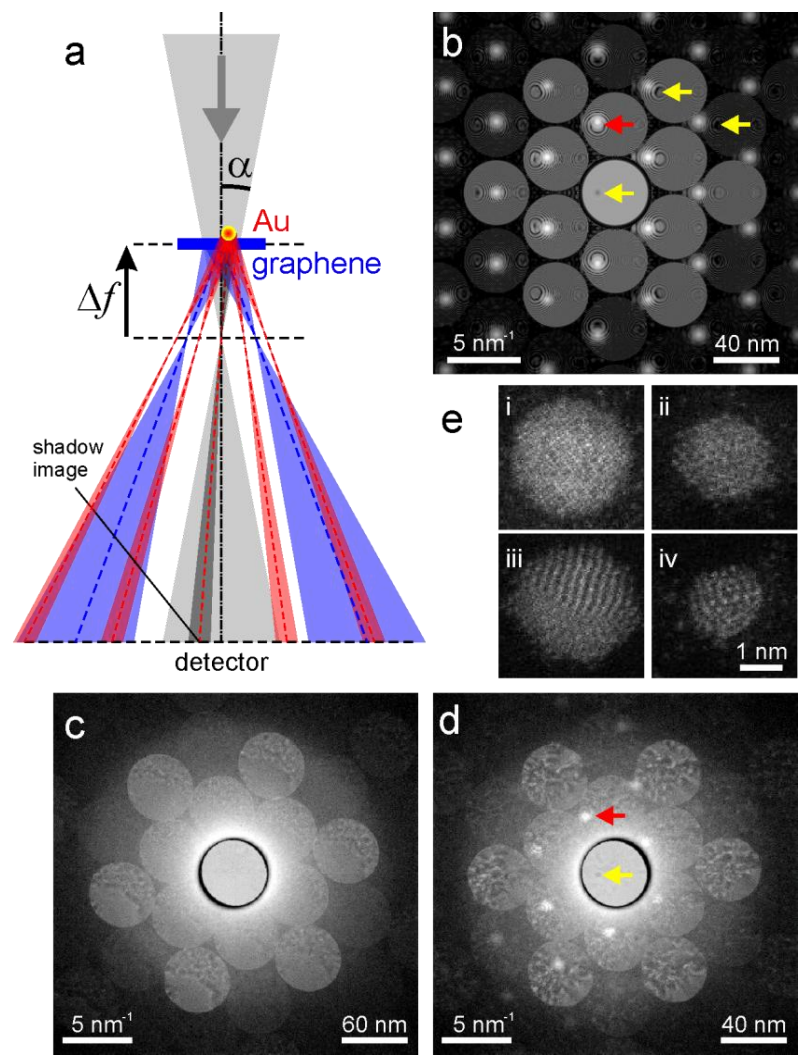

Fig. 1. Convergent beam electron diffraction (CBED) of Au nanoparticles on graphene. (a) Schematics of the experimental arrangement. The graphene sample with adsorbates is positioned above the focus point of the beam (underfocus); $\Delta f$ is the defocus, and $\alpha$ is the convergence semi-angle. (b) Simulated CBED pattern of graphene with a round Au nanoparticle of 4 nm in diameter, zone axis 011. (c) Experimental CBED pattern of graphene without Au nanoparticle, $\Delta f$ = -3 μm, the diameter of the probed area is about $D \approx 60$ nm. (d) Experimental CBED pattern of graphene with one Au nanoparticle in the probed region. In (b) and (d), $\Delta f$ = -2 μm, $D \approx 40$ nm. The red arrows in (b) and (d) indicate the CBED disks of the Au nanoparticles. The yellow arrows in (b) and (d) indicate the shadow images of the Au nanoparticle in the zero-order CBED disk (in-line hologram) and in the higher-order CBED disks. (e) High-angle annular dark-field (HAADF) images of Au nanoparticles deposited on graphene, showing different isomers: (i) round-shaped, (ii) icosahedron or cuboctahedron, (iii) and (iv) mixed-state Au nanoparticles. In (b), (c), and (d): $\alpha$ = 9.8 mrad, the left scalebars correspond to the detector plane, the right scalebars correspond to the sample plane, the intensity of the zero-order CBED spots is reduced by a factor of 1000 for the purpose of presentation, and the intensity is shown in logarithmic scale.

The explanation for why the zero-order signal of an Au nanoparticle is always bright in the diffraction pattern and is always dark (shadow image) in the CBED pattern can be provided by using the intensity conservation law as follows. In the absence of Au nanoparticles, the intensity in the zero-order CBED disk of graphene is uniform. When the probing wave illuminates the graphene with an Au nanoparticle on top, the signal diffracted by the graphene will be missing at the Au location. At the same time, the Au nanoparticle will diffract its own CBED pattern. The total intensity of the part of the probing wave that illuminates the Au nanoparticle is $I_0$. And thus, the total intensity of the Au nanoparticle's CBED pattern is also $I_0$. This gives an average intensity of $I_0/P$ for each of the $P$ CBED disks in the Au nanoparticle's CBED pattern. In total, at the location of the Au nanoparticle in the graphene's zero-order CBED disk, the intensity is reduced by the missing diffraction from the graphene ($-I_0$) and increased by diffraction from the Au nanoparticle ($+I_0/P$), thus always resulting in the intensity that is less than the intensity in the region of graphene. The Au nanoparticle always appears as a dark shadow image in the zero-order CBED disk of graphene.

The shadow images also appear in the higher-order CBED disks. They are formed as a result of the missing signal from the regions in the graphene support at the location of the Au nanoparticle. The carbon atoms in these regions do not contribute to the intensity in the graphene's CBED disks (Figs. 1a, b, and d). The shadow images of an Au nanoparticle appear at the same locations within graphene CBED disks – the location of the particle in the probed region. The shadow images can be seen in the simulated CBED patterns in Figs. 2, 3, 4, and 6, and in the experimental CBED patterns in Fig. 5.

## 2.2 Simulated CBED patterns

Only Au nanoparticles in a cuboctahedron isomer exhibit the FCC crystal structure; the corresponding CBED disks are observed in CBED patterns and can be easily indexed. For non-FCC structures of Au nanoparticles, the corresponding simulated diffraction and CBED patterns of an Au nanoparticle of 4 nm diameter in Dh and Ih isomers on graphene are shown in Fig. 2. The details of simulations are provided in Appendix 1.

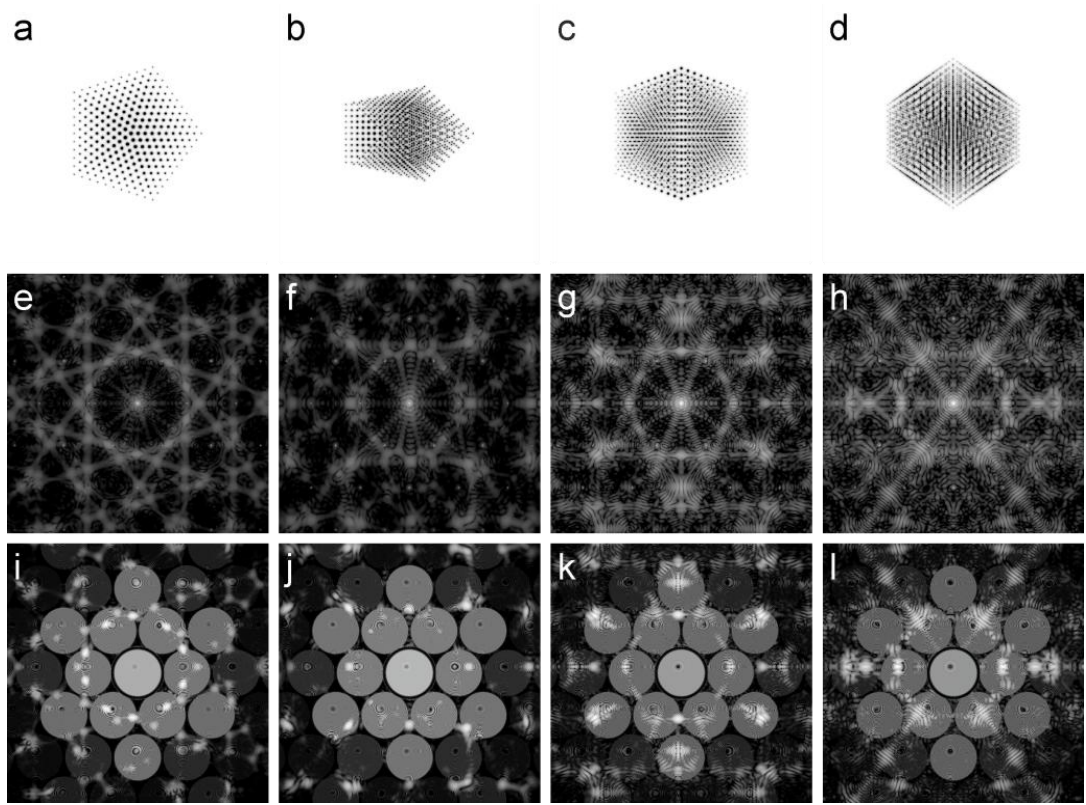

Fig. 2. Simulated diffraction and CBED patterns of an Au nanoparticle of 4 nm diameter in decahedron and icosahedron structures (non-FCC) on graphene. (a) – (b) decahedron structure when imaged at (a) 001 and (b) 011 zone axis. (c) – (d) icosahedron structure when imaged at (a) 001 and (b) 011 zone axis. (e) – (h) the corresponding diffraction patterns. (i) – (l) the corresponding CBED patterns, simulated at defocus $\Delta f$ = -2 μm, the convergence semi-angle $\alpha$ = 9.8 mrad; the zero-order CBED disk's intensity is reduced by factor 1000. In (e) – (l) the intensity is shown in logarithmic scale.

Simulated CBED patterns of a rotating Au nanoparticle (diameter of 4 nm) on graphene are shown in Fig. 3. The simulations demonstrate that as the nanoparticle rotates around its axis, the distribution of its CBED disks is changing, while the position and the contrast of the shadow image of the nanoparticle in the zero-order CBED disk (the in-line hologram of the Au nanoparticle) remain unchanged.

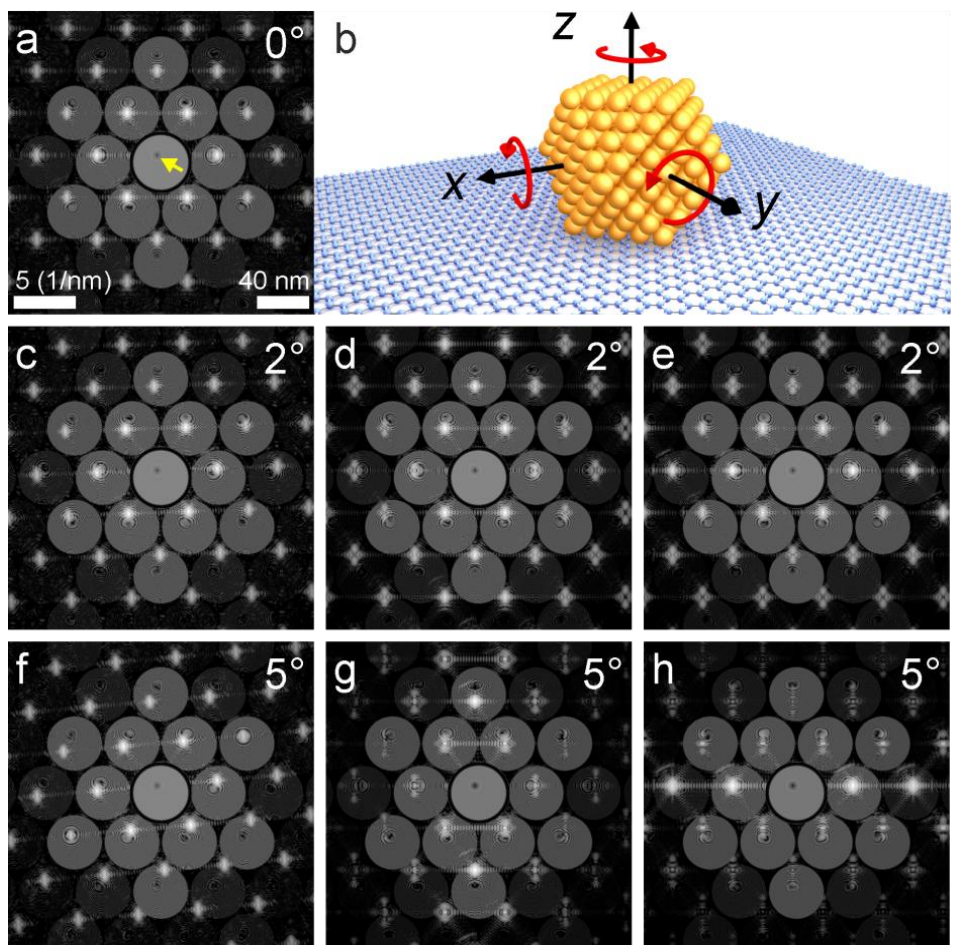

Fig. 3. Simulated CBED patterns of a rotating Au nanoparticle (4 nm diameter, cuboctahedron) on graphene. (a) CBED pattern of non-rotated (0°) particle; the yellow arrow indicates the position of the shadow image of the nanoparticle. (b) Illustration of the direction of the Au nanoparticle's rotation around the three axes. (c) - (h) The simulated CBED patterns for rotation around the $z$-axis ((c) and (f)), the y-axis ((d) and (g)), and the $x$-axis ((e) and (h)). The rotation angle is indicated in the top right corner. All the CBED patterns are shown in the same logarithmic intensity scale, and the intensity of the zero-order CBED spots was reduced by factor 2000 for presentation purposes. The parameters for simulations were selected to match the experimental

conditions: Au nanoparticle's orientation is [011], defocus $\Delta f$ = -2 μm, and the convergence semi-angle $\alpha$ = 9.8 mrad.

Figure 4 shows simulated CBED patterns of graphene with an additional graphene patch (mimicking contamination or adsorbate) and an Au nanoparticle. The intensity of the graphene patch is different in different CBED disks (Fig. 4), because it is given by both the stacking order and the *K*-coordinate of the CBED disk as $I \propto \cos(\vec{K}\Delta\vec{r})$, where $\vec{K} = (K_x, K_y)$ is the coordinate of the CBED disk, and $\Delta\vec{r} = (\Delta x, \Delta y, \Delta z)$ is the relative shift between the lattice of the graphene support and the lattice of the patch. Previously, the contrast in the CBED disks due to relative shifts between the two graphene layers was studied by theory and simulations in ref. [8]. A perfectly clean graphene monolayer exhibits CBED disks of featureless uniform intensity distributions. When such a perfect graphene support is laterally shifted, the positions and distributions of its CBED disk will remain the same, thus making it difficult to identify any shifts of the graphene. When an adsorbate or contamination is residing on the graphene, it creates "fingerprint" intensity variations – in-line holograms in graphene CBED disks. When graphene with the adsorbates is laterally shifted, these in-line holograms will also be shifted (Fig. 4a,b). Thus, the shifts of the graphene support with adsorbates can be determined from the shifts of the adsorbate's in-line holograms in the CBED disks. An Au nanoparticle creates its own CBED pattern. If the Au nanoparticle is shifting independently on the adsorbate, the nanoparticle's CBED pattern is also shifting (or rotating) independently on the CBED pattern of the adsorbate's in-line holograms (Fig. 4b,c). This way the movement of an Au nanoparticle can be separated from the movement of the adsorbate (support) – by inspecting the relative shifts of the corresponding CBED patterns. In addition, from the simulations shown in Fig. 4, it can be seen that some of the Au nanoparticle's CBED disks can overlap with the Au nanoparticle's shadow images, creating interference patterns.

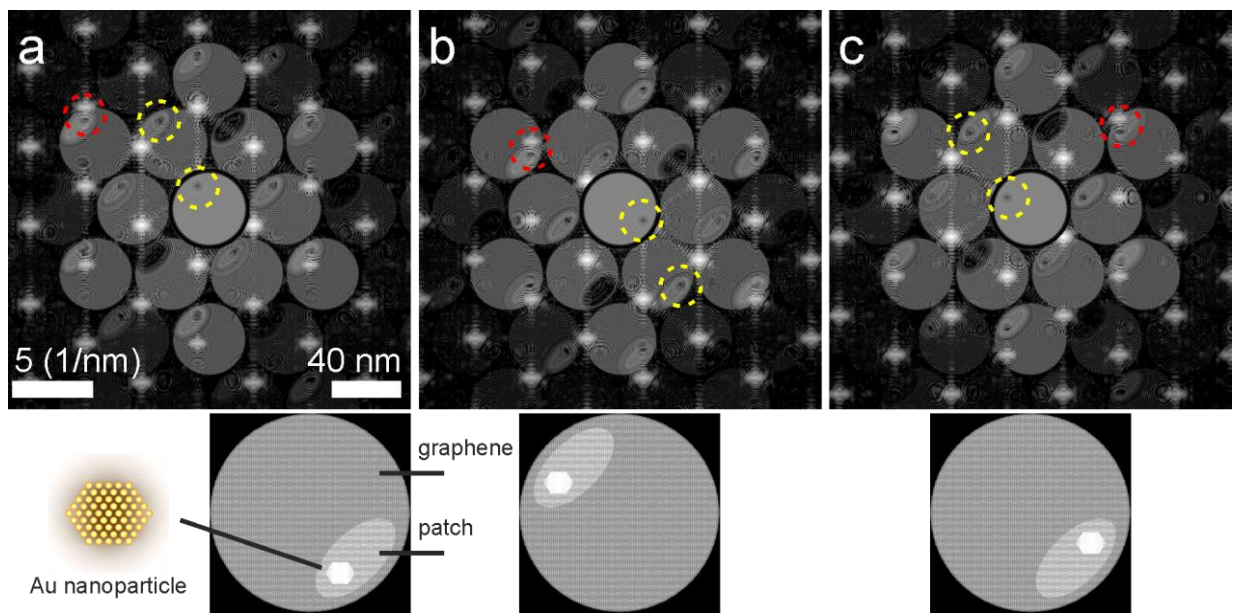

Fig. 4. Simulated CBED patterns of a 4 nm diameter cuboctahedron Au nanoparticle deposited on top of a second layer graphene patch on top of a graphene support. (a) – (c) CBED pattern of the samples. The insets below show the sample distribution; note that the distributions in the real-space and in a CBED disk are centro-symmetrically flipped. The yellow circles indicate some of the shadow images of the Au nanoparticle. The red circles indicate some of the regions where a shadow image of the Au nanoparticle and a CBED disk of the Au nanoparticle are in close proximity to each other and exhibit interference. Simulations are done at $\Delta f$ = -2 μm, and the convergence semi-angle $\alpha$ = 9.8 mrad. The CBED patterns are shown in logarithmic scale, and the intensity of the zero-order CBED spots was reduced by a factor of 2000 for presentation purposes.

## 3. Experimental results

For the CBED experiments sample, Au salt solution was drop-casted onto graphene (details are provided in Appendix 2), and crystalline Au particles of up to 5 nm in diameter were formed on the graphene surface (Fig. 1e). The electron microscopy experiments were performed using a Thermo Fisher Talos F200X S/TEM at 80 keV (equipped with Merlin direct electron detector) at a convergence semi-angle of 9.8 mrad that ensured maximal diameter but no overlapping between the CBED disks. No beam stopper was used, which allowed recording of the zero-order CBED disk. The magnification of the objective and condenser lenses was kept constant during acquisition, and the defocus was adjusted by moving the sample along the $z$ direction. The defocus was set to $\Delta f$ = -2 μm, thus providing the probed region of about 40 nm in diameter. The probe current was approx 50 pA; with the 40 nm diameter probed area, a 4 s exposure is approximately 10'000 e/Å². The probed region was not illuminated before the acquisition. The CBED patterns were collected during one session and under the same experimental conditions; for each region, a series of five frames was recorded with a 4 s exposure time per frame. An example of acquired CBED patterns is shown in Fig. 5. Time-series of CBED patterns for two selected regions are provided as SI Movies. In the CBED patterns, signals up to the 2$^{nd}$ and 3$^{rd}$ diffraction orders for graphene and up to the 7$^{th}$ order for Au were observed.

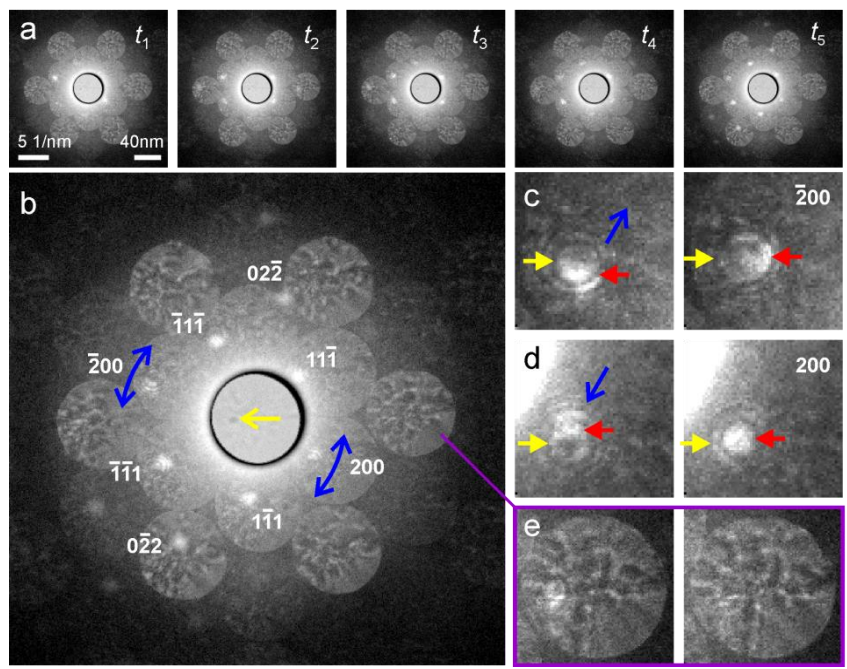

Fig. 5. Experimental CBED patterns of an Au nanoparticle rotating on graphene. (a) Time series of CBED patterns of the Au nanoparticle on graphene at $t_1$ = 0 s, $t_2$ = 4 s, $t_3$ = 8 s, $t_4$ = 12 s, and $t_5$ = 16 s; $\Delta f$ = -2 μm, the convergence semi-angle is $\alpha$ = 9.8 mrad, and the diameter of the probed area is about 40 nm. The left scalebar corresponds to reciprocal space (detector plane). The right scalebar corresponds to real space (sample plane). (b) Single-frame CBED pattern, the intensity of the zero-order CBED disk is reduced by a factor of 1000. The yellow arrow indicates the real-space position of the Au nanoparticle as a shadow image in the zero-order CBED disk. The blue arrows in (b) – (d) show the direction of the CBED disks' shifts during the time series acquisition. (c) and (d) are the same region in two CBED patterns from the time series, showing the shifts of the Au CBED disks' position in the direction indicated by the blue arrow. The red arrows indicate the location of the Au CBED disks, the yellow arrows indicate the location of the shadow images. (e) A selected CBED disk from two CBED patterns of the same time series showing a static distribution of the adsorbates signal, while an Au diffraction peak appears (left panel) and disappears (right panel).

## 3.1 Movements of Au nanoparticle

The exact movement of an Au nanoparticle on graphene can be retrieved from the presence of the following signal signatures in the CBED patterns. Since a CBED pattern is a defocused image of the sample, the lateral shifts of the Au nanoparticle on graphene result in the shift of the Au nanoparticle's CBED pattern within the CBED pattern of the entire sample (including the in-line hologram in the zero-order CBED disk), while the intensity of the Au nanoparticle's CBED disks does not change. When the Au nanoparticle is rotating along the *z*-axis (spinning), its CBED (diffraction) pattern is also rotating along the *z*-axis (spinning), and there is no change in the intensity of the CBED disks; the in-line hologram in the zero-order CBED disk is not shifting. When the Au nanoparticle is rotating around its *x*- or *y*-axis or rolling on the graphene, it is changing its orientation in the beam, and therefore its CBED disks can disappear and reappear at different positions.

Thus, the shifts and rotations of an Au nanoparticle can be well distinguished, and both can be picked up with high accuracy because of the atomic-resolution information available from the CBED disks' positions and intensity. The nanoparticle's CBED disks can disappear when the particle has melted or rotated so that no diffraction condition is fulfilled. To distinguish these two cases, the presence of the particle can be monitored from its shadow image in the zero-order CBED disk (in-line hologram). In the case of nanoparticle melting, its CBED disks disappear together with the shadow image in the zero-order CBED disk.

## 3.2 Rotation and rolling of Au nanoparticle

For each of the imaged Au nanoparticles, rich signal dynamics were observed from its CBED disks: the Au nanoparticle CBED disks changed their position and intensity relative to the CBED disks of graphene. For the Au nanoparticle shown in Fig. 5a, its diffraction pattern exhibited rotation around the nanoparticle's position (seen as a shadow image in the zero-order CBED disk in Fig. 5b). The rotation angle was evaluated from the Friedel's pair CBED disks indexed as -200 and 200. The CBED disks were moving in opposite directions, which is an indication of rotation rather than a lateral shift of the nanoparticle (Fig. 5c - d). The rotation was evaluated for each of the 5 frames relative to the first frame to be: 0°, -1.00°, -4.02°, +0.07°, +1.89° (where negative direction is counted clockwise), or the rotation between each consecutive frames is: 0°, -1.00°, -3.02°, +4.09°, +1.82° (details of the calculation are provided in Appendix 3). This back-and-forth rotation, or "jiggling", of the nanoparticle's CBED pattern is accompanied by a change in the CBED disks' intensities (a full sequence of frames is provided in Movie1) that can be explained by the particle re-orienting in the probing beam and, more accurately, by rolling back and forth on the graphene surface. The melting of the particle during the acquisition can be excluded because the last CBED pattern in the time series exhibited the maximal number of intense Au nanoparticles' CBED disks, and the shadow image of the Au nanoparticle was present in all CBED patterns in the time series.

In the experimental diffraction pattern shown in Fig. 5b, overlapping and interference between the shadow images and CBED disks of the Au nanoparticle can be observed. During the time series acquisition, the Au CBED disk (identified by the maximum of intensity) changes its position (due to the rotation of the nanoparticle), while the position of the shadow image (concentric rings with a darker center) remains almost unchanged (Fig. 5c - d).

The presence of adsorbates can help to separate the shift of the Au nanoparticle from the shifts of the contamination (adsorbates), as was discussed above and illustrated by the simulations shown in Fig. 4. For the Au nanoparticle shown in Fig. 5, the intensity distribution due to the adsorbates does not significantly change, neither in intensity nor in position, during the CBED time series acquisition, as can be seen in Fig. 5e. This means that the Au nanoparticle was moving on top of the adsorbates.

### 3.3 Facet diffraction lines

An interesting effect, which we name as "facet diffraction lines" was observed in some CBED patterns and is shown in Fig. 6. In the experimental CBED patterns, fast "jiggling" of the Au nanoparticle was accompanied by intense lines observed between the CBED disks; these lines were switching on and off (blinking), Fig. 6a – b and SI Movie2. Facet diffraction lines are observed when an intensity ray (formed by diffraction on a nanoparticle's facet) from one CBED disk coincides with an intensity ray from another CBED disk. The on-and-off behaviour of the facet diffraction lines could be explained by re-orientation of the Au nanoparticle; the simulations in Fig. 3g and h show that different orientation of the Au particle in the probing beam leads to the selected facet diffraction lines being turned on and off. The energy required to rotate an Au nanoparticle (estimated above) is small, which makes the rotational dynamics of the Au nanoparticle, and with this, the switching on and off of the facet diffraction lines, highly probable. Another explanation could be fast structural transformations between facetted (Ih, Dh) and round-shaped (Ch FCC) structural types. For a facetted Au nanoparticle, each CBED disk exhibits a ray of higher intensity in the direction orthogonal to the Au nanoparticle's facets (Fig. 6c), while a round-shaped FCC Au nanoparticle has CBED disks which exhibit concentric rings distribution without the rays (Fig. 6d). Preferred FCC structures and conversion of metastable Dh into FCC structures at room temperature were observed for $Au_{561}$ nanoparticles by aberration-corrected TEM studies [32].

## 3.4 Twinning

The CBED imaging mode allows distinguishing between several different particles and one particle with several different domains (twinning). For the same Au nanoparticle shown in Fig. 6, different sets of CBED disks were observed within the same CBED pattern. However, the real-space image of the sample available in the zero-order CBED disk showed the presence of only one Au nanoparticle (Fig. 6a,b). In addition, all of the Au CBED patterns that were observed in each individual experimental frame were centered at the same position – the shadow image of the Au particle in the zero-order CBED disk. This leads to the conclusion that all those Au CBED patterns can be attributed to a single Au particle whose position is seen as a dark feature in the zero-order CBED disk. The position of the Au nanoparticle remained unchanged during the time series acquisition.

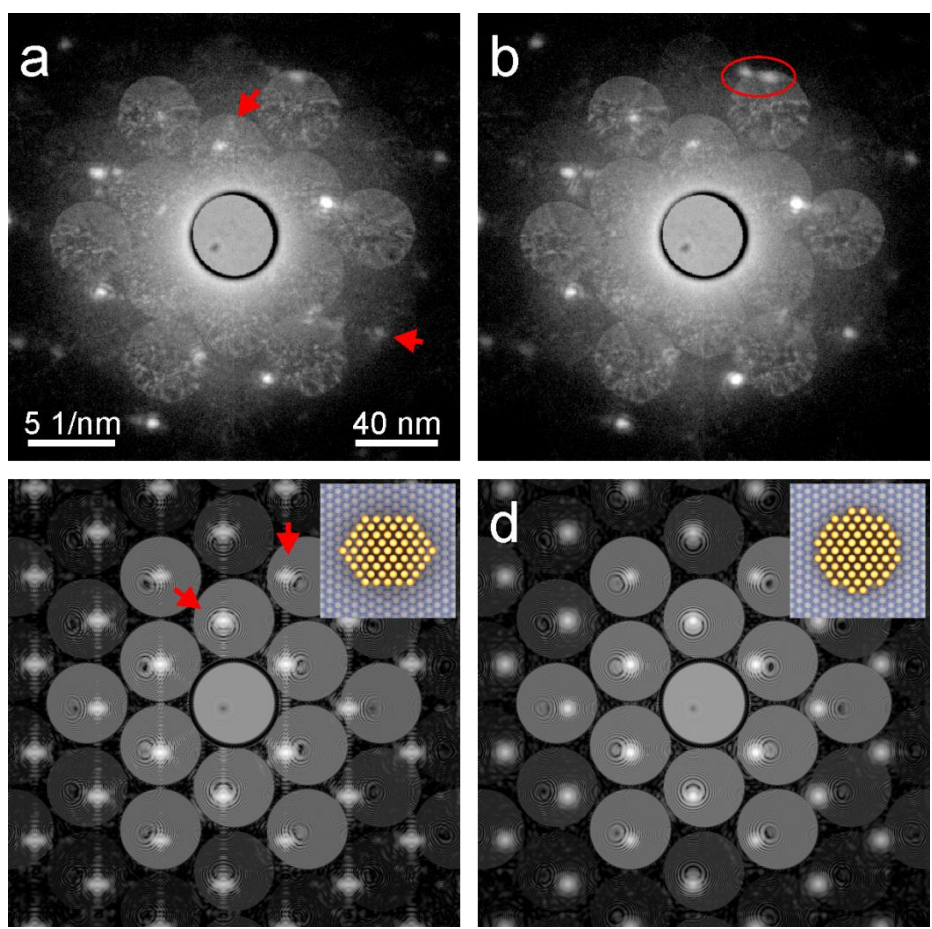

Fig. 6. Au nanoparticles' facet diffraction lines. (a) and (b) experimental CBED patterns from the same time series where the facet diffraction lines are switched on (a) and off (b). The red oval in (b) indicates the pair of CBED disks due to twinning. (c) and (d) simulated CBED patterns of a facetted cuboctahedron Au nanoparticle, zone axis 011 (c) and a round Au nanoparticle (d); the insets show the real-space distribution of Au

nanoparticles on graphene. The red arrows in (a) and (c) indicate the facet diffraction lines.

## 4. Discussion and Conclusions

In conclusion, CBED on 2D materials was employed for imaging tests object of relatively well known atomic structure and morphology – Au nanoparticles. It was observed that each Au nanoparticle provides its own CBED diffraction pattern superimposed with the CBED pattern of the supporting graphene. The CBED patterns of graphene and Au nanoparticles show independent behavior; thus changes in both the support and the nanoparticles could be separated. Each graphene CBED disk contains a shadow image of the Au nanoparticle that indicates the position of the nanoparticle in the probing beam. An interference pattern resulting from interaction between the CBED disks and shadow images of the Au nanoparticle was observed.

By using the CBED technique, the real-space image of the Au nanoparticle on graphene and the diffraction pattern of the same probed region were recorded in one single-shot intensity measurement. During the time series CBED patterns acquisition, the Au nanoparticle's CBED disks exhibited changes in their positions and intensities, and even the full disappearance of the CBED disks. The presence of Au nanoparticles in the probing electron beam was monitored by the shadow images of the Au nanoparticles in the zero-order CBED disk. Combining the real- and diffraction-space information allowed reconstruction of the dynamic behavior of the nanoparticle – rolling back and forth on the graphene surface. The estimated energies required for the nanoparticle's lateral and rotational movements are very small, and thus these movements are highly probable at room temperature even without additional energy deposited by the electron beam. It was observed that Au nanoparticles were most commonly located in the contaminated areas, which agrees with previous observations [46]. These two effects combined can provide an explanation for the particles' back-and-forth rolling: the nanoparticle could be moving within a confined adsorbate region that provided an energetically favorable support. Recent studies suggest that rapid motion and coalescence of Au nanoparticles during TEM studies can be caused by the charge built up in Au nanoparticles due to the presence of a non-conducting support [51].

The zero-order CBED disk is an in-line hologram of the sample, and the sample distribution could be in principle directly reconstructed by applying in-line holography reconstruction algorithms [15, 17]. However, in the observed CBED patterns, the shadow images of Au nanoparticles in the zero-order disks are blurred due to fast movements of the nanoparticles; they do not exhibit interference fringes characteristic for an in-line hologram and therefore cannot be reconstructed by applying the holography algorithms. Yet, in the approach demonstrated here, the diameter and position of an Au nanoparticle can be evaluated even from a blurred shadow image: the diameter of the particle shown in Fig. 5 is about 4.3 nm. Due to fast movements and rotations of the Au nanoparticle that led to blurred interference patterns in individual CBED disks, the entire CBED patterns shown in this study cannot be used for reconstruction by iterative phase retrieval. However, the blurriness of the zero-order CBED disk cannot be stated as a disadvantage of the CBED technique, because it results from the fast dynamics of the Au nanoparticles on the graphene support, and is not a limitation of the CBED technique itself. Atomic-resolution reconstruction, in principle, should be possible from a CBED pattern of a static object or from a CBED pattern of a dynamic object, provided that this CBED pattern is acquired within a sufficiently short time interval relative to the dynamic motion of the object.

For dynamic objects, to quantify how fast data can be acquired while still obtaining good signal, we made the following estimations. To obtain a certain number of counts per pixel (cpp) in the zero-order CBED disk, the acquisition time is given by $\Delta t = \mathrm{cpp} S_{zo} e / j$, where $S_{zo}$ (in pixels) is the area occupied by the zero-order disk in the detector plane (the area depends on the selected aperture's diameter), $e$ is the elementary charge and $j$ is the current. In order to get a signal with SNR = 5, according to the Rose criterion, the cpp value should be at least 25. For the parameters that were used in our experiments, $j$ = 50 pA, and the diameter of the zero-order CBED disk of 92 pixels, we obtain $\Delta t$ = 0.53 ms. This is the exposure time limit to obtain a signal with SNR = 5 in the zero-order CBED disk. At 80 keV electron energy, the intensity in a first-order CBED disk is typically less than the intensity in the zero-order CBED disk by a factor of 1000; the ratio of the zero-order to the first-order intensities depends on the electron energy and is only 25 – 50 for low-energy electrons of 120 – 230 eV energy [19, 52]. Thus for 80 keV electron energy, to obtain a signal with SNR = 5 in a first-order CBED disk, the exposure time should be at least 0.5 s. The exposure time can also be reduced by using a higher current.

In this study, three individual Au nanoparticles were studied in detail. For each of the three Au nanoparticles, the diameter was evaluated from the shadow image in the zero-order CBED disk, giving 4.7, 4.3 and 4.3 nm, respectively. Overall, we observed about 50 Au nanoparticles with diameters of approximately 4 nm and larger that exhibited CBED patterns. This size limit could be due to the fact that particles of smaller sizes do not have FCC structure and only an FCC produces a distinct diffraction pattern. This feature can be used for monitoring the structural transformations: if a particle exists in a non-FCC structure, its presence should be visible as shadow images. Should the particle transform into an FCC structure, Au CBED disks should appear in addition to the shadow images.

In addition, CBED on 2D materials is potentially promising for imaging biological macromolecules deposited on graphene [53]. The radiation dose can be regulated by selecting the defocus distance [54]. The radiation dose required to record a good signal-to-noise ratio hologram is generally much less than the radiation dose required for recording a diffraction pattern that exhibits a few counts per pixel at the detector's edge [55]. The sample can be reconstructed from the entire CBED pattern (with the resulting resolution given by the highest order of the detected signal) by using iterative phase retrieval algorithms [18, 56, 57] or from an individual CBED disk (although at lower resolution given by the CBED disk's diameter) by using holography reconstruction algorithms [17, 18]. The interference patterns in the individual CBED disks are the in-line holograms formed by illuminating the sample with the wave diffracted from the support (graphene) at different angles. Thus, these in-line holograms are analogous to tomographic measurements and contain 3D information. The principles of CBED pattern formation are the same as for divergent probing wavefront, and they can be applied for low-energy electron holography of macromolecules deposited onto graphene – a relatively novel technique for imaging individual proteins [19, 52, 58, 59]. For low-energy electrons, their relatively large wavelengths (0.123 nm for 100 eV electrons) result in graphene support diffracting at large angles (first-order diffraction at 35° for 100 eV electrons) and thus probing the macromolecule at large tilt angles, almost like in a tomography experiment.

Recently, a very similar imaging scheme was demonstrated by using x-rays, where the convergent wavefront was modulated by a periodical grid (similar to graphene support in CBED) before the probing an object. Here, the technique was named "single-shot ptychography" because of the overlapping beams of light arranged in a grid pattern simultaneously illuminating a sample, allowing a full ptychographic dataset in a single shot [60].

## Appendix 1: Simulations

**Atomic structures:** The coordinates of Au atoms in an Au nanoparticle were generated using the MATLAB code "Cluster Generator" [61]. The atomic coordinates were rotated to achieve the desired orientation of the Au nanoparticle in the probing beam by applying rotational matrix transformations.

**Transmission function of graphene**: The input data consisted of an array of the coordinates of atoms $(x_n, y_n)$ in the graphene monolayer. The transmission function of a monolayer graphene was calculated as:

$$t(x,y) = \exp\left[i\sigma v_z(x,y) \otimes l(x,y)\right],$$

where $v_z(x,y)$ is the projected potential of an individual atom, $l(x,y)$ is the function describing positions of the atoms in the lattice, $\otimes$ denotes convolution, and $\sigma$ is the interaction parameter; $\sigma = 2\pi m e \lambda / h^2$, where $m$ is the relativistic mass of electron, $e$ is an elementary charge, $\lambda$ is wavelength of electrons, and *h* is Planck's constant. The projected potential of a single carbon atom was simulated in the form (ref. [62]):

$$v_z(r) = 4\pi^2 a_0 e \sum_{i=1}^{3} a_i K_0\left(2\pi r \sqrt{b_i}\right) + 2\pi^2 a_0 e \sum_{i=1}^{3} \frac{c_i}{d_i} \exp\left(-\pi^2 r^2 / d_i\right),$$

where $r = \sqrt{x^2 + y^2}$, $a_0$ is the Bohr radius, $K_0(...)$ is the modified Bessel function, and $a_i, b_i, c_i, d_i$ are parameters that are provided in ref [62]. In $v_z(r)$, the singularity at $r=0$ is replaced by the value of $v_z(r)$ at $r$ = 0.1 Å. The convolution in $t(x,y)$ was calculated as $\text{FT}^{-1}\{\text{FT}[v_z(x,y)]\text{FT}[l(x,y)]\}$, where FT and FT$^{-1}$ mean Fourier and inverse Fourier transform, respectively. $\text{FT}[l(x,y)]$ was simulated without applying fast Fourier transform (FFT) in order to avoid artefacts associated with FFT as $\text{FT}[l(x,y)] = \sum_n \exp\left[-i\left(k_x x_n + k_y y_n\right)\right]$, where $(x_n, y_n)$ are the exact atomic coordinates (not pixels) and $(k_x, k_y)$ are the coordinates in the Fourier plane. $\text{FT}^{-1}\{\text{FT}[v_z(x,y)]\text{FT}[l(x,y)]\}$ was calculated by applying an inverse FFT to the product of $\text{FT}[v_z(x,y)]$ and $\text{FT}[l(x,y)]$. The monolayer graphene was simulated for a region of 80 × 80 nm$^2$, sampled with 5634 × 5634 pixels (the largest numbers at which it was possible to run the simulations due to computational constraints).

**Transmission function of Au nanoparticle:** The atomic coordinates of a 3D Au nanoparticle were projected onto one plane. The transmission function of the Au nanoparticle in this plane was simulated using the same protocol as the transmission function of graphene, with the projected potentials of Au atoms.

**Transmission function of graphene with an Au nanoparticle** was calculated as a product of the transmission function of graphene and transmission function of Au nanoparticle.

**Probing electron beam**: For CBED, the complex-valued probing electron beam was simulated as explained in detail in ref. [21]. For diffraction patterns, the probing electron beam was calculated as a uniform distribution limited by a round aperture with blurred edges (the edges were blurred by applying an apodization cosine-filter [17]).

**CBED and diffraction patterns**: The exit wave was calculated as a product of the complex-valued probing wave and the complex-valued transmission function of the sample. The wavefront distribution in the detector plane was obtained by calculating FFT of the exit wave. The intensity distribution was obtained by calculating square of the absolute value of the wave wavefront distribution in the detector plane. The intensity distributions sampled with 5634 × 5634 pixels were then cropped to the size of 2000 × 2000 pixels.

## Appendix 2: Sample preparation

Au nanoparticles were deposited on suspended graphene at the National Graphene Institute in Manchester, UK, using the following protocol: a foil of chemical vapour deposited (CVD) on copper graphene was spin coated with PMMA in anisole, baked and plasma-cleaned on the backside of the copper foil. The copper was removed by chemical etching, and the graphene layer was transferred onto the SiN TEM grid. PMMA layer was removed by hot solvent cleaning. Solution of Au salt ($HAuCl_4 \cdot 3H_2O$, Au(III)-chloride trihydrate) was drop-casted onto the graphene coated grids and let to dry. Using this protocol, Au nanoparticles of different diameters were produced, ranging from atomically dispersed adatoms to nanoparticles of around 5 nm in diameter. High-angle annular dark-field (HAADF) images of the sample is shown in Fig. A1.

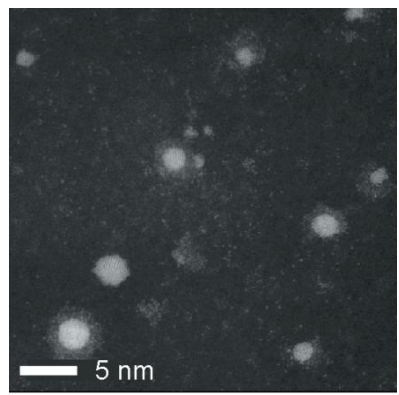

Fig. A1. High-angle annular dark-field (HAADF) image of Au nanoparticles deposited on graphene. The images were acquired with 200 keV energy electrons using a Titan G2 80 – 200keV S/TEM.

## Appendix 3: Evaluation of rotation of Au nanoparticle 's CBED pattern

Each of the five CBED patterns in the time sequence exhibited a different number of Au nanoparticle's CBED disks, but two CBED disks were present in all of the CBED patterns (indicated in Fig. 5): peak -200 (peak1) and peak 200 (peak2). The positions of these two CBED disks $\left(x_i^{(1)}, y_i^{(1)}\right)$ and $\left(x_i^{(2)}, y_i^{(2)}\right)$, and the position of the shadow image of the crystal in the zero-order CBED disk $\left(x_i^{(0)}, y_i^{(0)}\right)$, where *i* = 1…5 is the frame number, were extracted as the maxima of the Gaussian fit of the intensity (minimum in the shadow image). The position of the shadow image is the origin of the Au nanoparticle's CBED pattern, and the relative positions of the CBED disks to the origin at $\left(x_i^{(0)}, y_i^{(0)}\right)$ were calculated for each CBED pattern in the time series.

To compare the rotation between CBED pattern 1 and CBED pattern *i*, the coordinates of the CBED disks' positions in CBED pattern 1 were rotated around the Au nanoparticle's position. The rotation of the coordinates was calculated using the rotation matrix:

$$x_{\text{rot}} = x\cos\varphi + y\sin\varphi.$$
$$y_{\text{rot}} = x\sin\varphi - y\cos\varphi$$

A series of rotations was calculated with a step of $\Delta\varphi = 0.01°$. For each rotation angle, the matching between the rotated CBED disks' positions and the CBED disks' positions in the CBED pattern *i* was evaluated by calculating the standard deviation as:

$$\text{SD} = \sqrt{\frac{1}{4}\left[\left(x_{1,\text{rot}}^{(1)} - x_i^{(1)}\right)^2 + \left(y_{1,\text{rot}}^{(1)} - y_i^{(1)}\right)^2 + \left(x_{1,\text{rot}}^{(2)} - x_i^{(2)}\right)^2 + \left(y_{1,\text{rot}}^{(2)} - y_i^{(2)}\right)^2\right]}.$$

The rotation angle which provided the least standard deviation was identified as the correct rotation angle.

## CRediT authorship contribution statement

## Declaration of competing interest
The authors declare they have no known competing financial interest or personal relationships that could have appeared to influence the work reported in this paper.

## Data availability
The authors declare no competing financial interest.

## Acknowledgements
SM, DP and TL acknowledge Swiss National Science Foundation (SNSF) Research Grant 200021_197107. KSN acknowledges support from the Ministry of Education, Singapore (Research Centre of Excellence award to the Institute for Functional Intelligent Materials, I-FIM, project No. EDUNC-33-18-279-V12), the National Research Foundation, Singapore under its AI Singapore Programme (AISG Award No: AISG3-RP-2022-028) and from the Royal Society (UK, grant number RSRP\R\190000). SJH and ML thank the Engineering and Physical Sciences Research Council (EPSRC) for funding under grants EP/Y024303, EP/S021531/1, EP/M010619/1, EP/V007033/1, EP/S030719/1,

EP/V001914/1, EP/V036343/1 and EP/P009050/1 and the European Research Council (ERC) under the European Union's Horizon 2020 research and innovation programme (Grant ERC-2016-STG-EvoluTEM-715502 and QTWIST (no. 101001515). TEM access was supported by the Henry Royce Institute for Advanced Materials, funded through EPSRC grants EP/R00661X/1, EP/S019367/1, EP/P025021/1 and EP/P025498/1.